\documentclass[twocolumn,showpacs,preprintnumbers,superscriptaddress,amsmath,amssymb]{revtex4}
\usepackage{amssymb}
\usepackage{graphicx}
\usepackage{array}
\usepackage{hhline}
\usepackage{longtable}

\renewcommand{\thetable}{\Roman{table}} \thetable

\allowdisplaybreaks[1]

\begin{document}

\title{Different thresholds of bond percolation in scale-free networks \\with identical degree sequence}

\author{Zhongzhi Zhang}
\email{zhangzz@fudan.edu.cn}

\author{Shuigeng Zhou}
\email{sgzhou@fudan.edu.cn}

\author{Tao Zou}

\author{Lichao Chen}

 \affiliation{School of Computer Science, Fudan University,
Shanghai 200433, China}%
 \affiliation{Shanghai Key Lab of
Intelligent Information Processing, Fudan University, Shanghai
200433, China}

\author{Jihong Guan}
\email{jhguan@tongji.edu.cn}

\affiliation{Department of Computer Science and Technology, Tongji
University, 4800 Cao'an Road, Shanghai 201804, China}

\date{\today}

\begin{abstract}

Generally, the threshold of percolation in complex networks depends
on the underlying structural characterization. However, what
topological property plays a predominant role is still unknown,
despite the speculation of some authors that degree distribution is
a key ingredient. The purpose of this paper is to show that
power-law degree distribution itself is not sufficient to
characterize the threshold of bond percolation in scale-free
networks. To achieve this goal, we first propose a family of
scale-free networks with the same degree sequence and obtain by
analytical or numerical means several topological features of the
networks. Then, by making use of the renormalization group technique
we determine the threshold of bond percolation in our networks. We
find an existence of non-zero thresholds and demonstrate that these
thresholds can be quite different, which implies that power-law
degree distribution does not suffice to characterize the percolation
threshold in scale-free networks.

\end{abstract}

\pacs{05.70.Fh, 89.75.Hc, 64.60.Ak, 87.23.Ge}

%05.45.Df  Fractals
%47.53.+n  Fractals
%89.75.Da  Systems obeying
%scaling laws 89.75.Fb Structures and organization in complex systems
%64.60.Ak Renormalization-group, fractal, and percolation studies of phase transitions
%89.75.Da Systems obeying scaling laws
%89.75.Fb Structures and organization in complex systems
%89.75.Hc Networks and genealogical trees
%89.75.-k Complex systems
%05.10.-a Computational methods in statistical physics and nonlinear
%                dynamics
%36.40.Qv Stability and fragmentation of clusters
%87.19.Xx Diseases
%87.23.Ge Dynamics of social systems
%05.70.Fh Phase transitions: general studies

 \maketitle

\section{Introduction}

As one of the best studied problems in statistical physics,
percolation~\cite{StAh92} is nowadays also the subject of intense
research in the field of complex networks~\cite{AlBa02}. In a
network, if a fraction of its vertices (nodes, sites) or edges
(links, bonds) is chosen independently with a probability $p$ to be
``occupied", it may undergo a percolation phase transition: when $p$
is above a threshold value $p_c$, called percolation threshold, the
network possesses a giant component consisting of a finite fraction
of interconnected nodes; otherwise, the giant component disappears
and all nodes disintegrate into small clusters. So far, percolation
in complex networks has received considerable attention in the
community of statistical physics~\cite{DoGoMe08}, because it is not
only of high theoretical interest, but also relevant to many aspects
of networks, including network
security~\cite{AlJeBa00,CoErAvHa00,CaNeStWa00,CoErAvHa01}, disease
spread on networks~\cite{MoNe00,Ne02,BoVePa03,Da05,ZhZhZoCh08}, etc.

Since global physical properties of random media alter substantially
at the percolation threshold, which is central to understanding and
applying this process, thus the precise knowledge of percolation
threshold is extremely important~\cite{ScZf08}. The issue of
determining or calculating the percolation threshold has been the
subject of intense study since the introduction of the model over
half a century ago~\cite{BrHa57,Fl41}. Despite decades of effort,
there is still no general method for computing the percolation
threshold of arbitrary graphs, and rigorous solution for percolation
threshold is confined to some special
cases~\cite{ScZf08,Sc06,Zf06,Pi06}, such as the Barab\'asi-Albert
(BA) network~\cite{BaAl99}, two dimensional lattice, and some other
lattices. In most cases (e.g. lattices in three dimensions or
above), the percolation threshold is estimated with numerical
simulations, which are often time-consuming~\cite{NeZi01}. Thus,
finding the threshold exactly is essential to investigating the
percolation problem on a particular graph~\cite{Zf06}.

Perhaps the main reason for studying percolation in complex networks
is to understand how the percolation properties are influenced by
underlying topological structure. It has been established that
degree distribution has a qualitative impact on the percolation.
Recent studies indicated that in uncorrelated scale-free networks
the percolation threshold is absent~\cite{CoErAvHa00,CaNeStWa00}.
Then a lot of other jobs followed, studying the influences of other
properties on the percolation properties in scale-free networks;
these include degree correlations~\cite{BoVePa03,VaMo03}, clustering
coefficient~\cite{SeBo06}, and so forth. It was found that, degree
correlations and clustering coefficient can strongly affect some
percolation properties, but they cannot restore a finite percolation
threshold in scale-free networks. This raises the question as to
whether scale-free degree distribution is the only ingredient
responsible for the absence of the percolation threshold. In other
words, whether power-law degree distribution suffices to
characterize the zero percolation threshold in scale-free networks.

In this paper, we study the effects of power-law degree distribution
on the percolation threshold in scale-free networks. To this end, we
first construct a class of scale-free networks with identical degree
sequence by introducing a control parameter $q$. We then study
analytically or numerically the topological features of the networks
and show that this class of networks has unique topologies. Finally,
using the renormalization-group theory, we investigate analytically
the bond percolation problem in the considered networks, and find
the existence of non-zero percolation thresholds depending on
parameter $q$. Our findings indicate that the degree distribution by
itself is not enough to characterize the percolation thresholds in
scale-free networks. On the other hand, since our networks have the
same degree sequence and thus the same degree distribution, the
model proposed here can serve as a useful tool (substrate model) to
check the impact of power-law degree distribution on the dynamical
processes taking place on top of scale-free networks.

\section{Network construction and structural characteristics}

In this section, we study the construction and structural properties
of the networks under consideration, with focus on degree
distribution, clustering coefficient, and average path length (APL).

%%%%%%%%%%%%%%%%%%%%%%%%%%%%%%%%%%%%%%%%%%%%%%%%%%%%%%%%%
% Figure 1
%%%%%%%%%%%%%%%%%%%%%%%%%%%%%%%%%%%%%%%%%%%%%%%%%%%%%%%%%%
\begin{figure}[h]
\includegraphics[width=0.6\linewidth,trim=100 0 100 10]{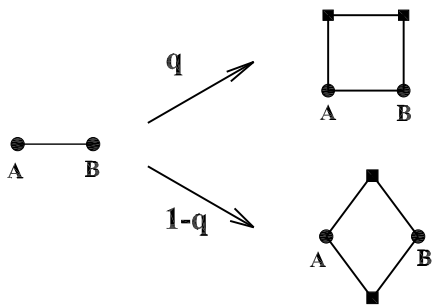}
\caption{Iterative construction method of the
networks. Each link is replaced by either of the connected clusters
on the right-hand side of arrows with a certain probability, where
black squares represent new vertices.}\label{fig1}
\end{figure}
%%%%%%%%%%%%%%%%%%%%%%%%%%%%%%%%%%%%%%%%%%%%%%%%%%%%%%%%%%

\subsection{Construction algorithm}

\begin{figure}
\begin{center}
\includegraphics[width=0.90\linewidth,trim=140 0 100 5]{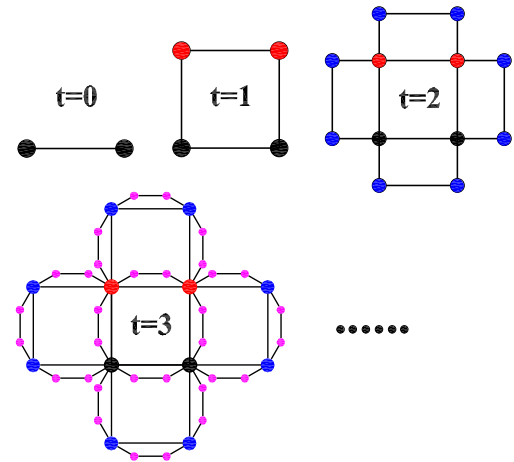}
\caption{(Color online) Illustration of the first four evolution
steps of the network growth process for the particular case $q=1$.}
\label{flower}
\end{center}
\end{figure}

\begin{figure}[h]
\centering
\includegraphics[width=0.75\linewidth,trim=110 0 100 20]{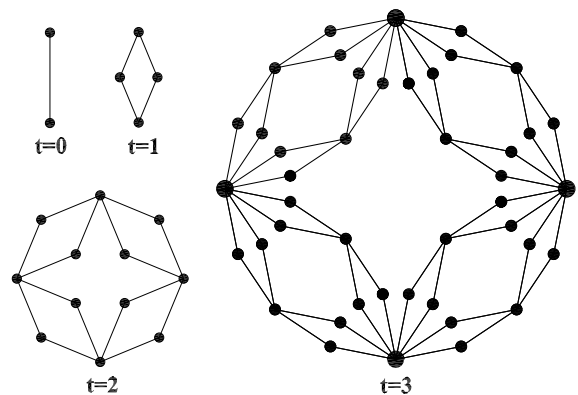}
\caption{Sketch of the iteration process of the network
for the particular case of $q=0$.}\label{fractal}
\end{figure}

The proposed networks (graphs) are constructed in an iterative way
as shown in Fig.~\ref{fig1}. Let $H_{t}$ ($t\geq 0$) denote the
networks after $t$ iterations. Then the networks are built in the
following way: for $t=0$, the initial network $H_{0}$ is two nodes
connected by an edge. For $t\geq 1$, $H_{t}$ is obtained from
$H_{t-1}$. We replace each existing link in $H_{t-1}$ either by a
connected cluster of links on the top right of Fig.~\ref{fig1} with
probability $q$, or by the connected cluster on the bottom right
with complementary probability $1-q$. The growing process is
repeated $t$ times, with the graphs obtained in the limit $t \to
\infty$. Figures~\ref{flower} and~\ref{fractal} show the growth
process of two networks for two limiting cases of $q=0$ and $q=1$,
respectively.

%%%%%%%%%%%%%%%%%%%%%%%%%%%%%%%%%%%%%%%%%%%%%%%%%%%%%%%%%
% Figure 2
%%%%%%%%%%%%%%%%%%%%%%%%%%%%%%%%%%%%%%%%%%%%%%%%%%%%%%%%%%
%%%%%%%%%%%%%%%%%%%%%%%%%%%%%%%%%%%%%%%%%%%%%%%%%%%%%%%%%%

%%%%%%%%%%%%%%%%%%%%%%%%%%%%%%%%%%%%%%%%%%%%%%%%%%%%%%%%%
% Figure 3
%%%%%%%%%%%%%%%%%%%%%%%%%%%%%%%%%%%%%%%%%%%%%%%%%%%%%%%%%%
%%%%%%%%%%%%%%%%%%%%%%%%%%%%%%%%%%%%%%%%%%%%%%%%%%%%%%%%%%

Now we compute some related quantities such as the number of total
nodes and edges in $H_t$, called network order and size,
respectively. Let $L_v(t)$ be the number of nodes generated at step
$t$, and $E_t$ the total number of edges present at step $t$. Then
$L_v(0)=2$ and $E_0=1$. By construction (see Fig.~\ref{fig1}), we
have $E_t=4E_{t-1}=4^t$ ($t\geq 0$). On the other hand, each
existing edge at a given step will yield two new nodes at next step,
this leads to $L_v(t)=2E_{t-1}=2\times4^{t-1}$ ($t\geq 1$). Then the
number of total nodes $N_t$ present at step $t$ is
\begin{equation}\label{Nt1}
N_t=\sum_{t_i=0}^{t}L_v(t_i)=\frac{2}{3}(4^t+2).
\end{equation}
The average node degree after $t$ iterations is $\langle k
 \rangle_t=\frac{2\,E_t}{N_t}=\frac{3\times4^t}{4^t+2}$,
which approaches 3 for large $t$.

\subsection{Degree distribution}

When a new node $i$ is added to the networks at a certain step $t_i$
($t_i\geq 1$), it has a degree of 2. We denote by $k_i(t)$ the
degree of node $i$ at time $t$. By construction, the degree $k_i(t)$
evolves with time as $k_i(t)=2\,k_i(t-1)=2^{t+1-t_i}$. That is to
say, the degree of node $i$ increases by a factor 2 at each time
step. Thus, the degree spectrum of the networks is discrete. In
$H_t$ all possible degrees of nodes is 2, $2^2$ $2^3$, $\ldots$,
$2^{t-1}$, $2^t$; and the number of nodes with degree $k=2^{t+1-m}$
is $n_k=L_v(m)=4^{m-1}$. Therefore, all class of networks $H_t$ have
the same degree sequence (thus the same degree distribution) in the
full range of $q$.

Since the degree spectrum of the networks is not continuous, it
follows that the cumulative degree distribution~\cite{Ne03} is given
by $P_{\rm cum}(k)=\frac{N_{t,k}}{N_t}$, where $N_{t,k}=\sum_{k'\geq
k}n_{k'}$ is the number of nodes whose degree is not less than $k$.
When $t$ is large enough, we find $P_{\rm cum}(k)\sim k^{-2}$. So
the degree distribution $P(k)$ of the networks follows a power-law
form $P(k)\sim k^{-\gamma}$ with the exponent $\gamma=3$,
independent of $q$. Notice that the same degree exponent has been
obtained in the famous BA network~\cite{BaAl99}.

\subsection{Clustering Coefficient}

By definition, the clustering coefficient~\cite{WaSt98} of a node
$i$ with degree $k_i$ is given by $C_i =2e_i/[k_i(k_i-1)]$, where
$e_i$ is the number of existing triangles attached to node $i$, and
$k_i(k_i-1)/2$ is the total number of possible triangles including
$i$. The clustering coefficient of the whole network is the average
over all individual $C_i's$. By construction, there are no triangles
in $H_t$, so the clustering coefficient of every node and their
average value in $H_t$ are both zero.

\subsection{Average path length}

Let $d_{ij}$ represent the shortest path length from node $i$ to
$j$, then the average path length $d_{t}$ of $H_t$ is defined as the
mean of $d_{ij}$ over all couples of nodes in the network:
\begin{equation}\label{APL01}
  d_{t} = \frac{D_t}{N_t(N_t-1)/2}\,,
\end{equation}
where
\begin{equation}\label{APL02}
  D_t =\sum _{\substack{i \in H_t,\, j \in  H_t\\ i \neq j}} d_{ij}
\end{equation}

denotes the sum of the shortest path length between two nodes over
all pairs.

For general $q$, it is difficult to derive a closed formula for the
APL $d_{t}$ of $H_t$. But for two limiting cases of $q=1$ and $q=0$,
both the networks are deterministic, we can obtain the analytic
solutions for APL.

%%%%%%%%%%%%%%%%%%%%%%%%%%%%%%%%%%%%%%%%%%%%%%%%%%%%%%%%%
% Figure 4
%%%%%%%%%%%%%%%%%%%%%%%%%%%%%%%%%%%%%%%%%%%%%%%%%%%%%%%%%%
\begin{figure}
\begin{center}
\includegraphics[width=.8\linewidth,trim=100 0 100 0]{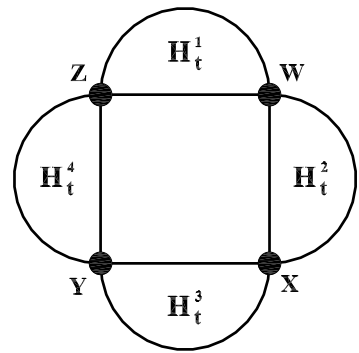}
\caption{Second construction method of the network for $q=1$ case
that highlights self-similarity: The graph after $t+1$ construction
steps, $H_{t+1}$, is composed of four copies of $H_t$ denoted as
$H_t^{\theta}$ $(\theta=1,2,3,4)$, which are connected to one
another as above.} \label{copy}
\end{center}
\end{figure}
%%%%%%%%%%%%%%%%%%%%%%%%%%%%%%%%%%%%%%%%%%%%%%%%%%%%%%%%%%

\subsubsection{Case of $q=1$}

In the special case (see Fig.~\ref{flower}), the networks are
reduced to the $(1,3)$-flower proposed in~\cite{RoHaAv07}. This
limiting case of network has a self-similar structure that allows
one to calculate $d_{t}$ analytically. The self-similar structure is
obvious from an equivalent network construction method: to obtain
$H_{t+1}$, one can make four copies of $H_t$ and join them in the
hub nodes. As shown in Fig.~\ref{copy}, network $H_{t+1}$ may be
obtained by the juxtaposition of four copies of $H_t$, which are
consecutively labeled as $H_{t}^{1}$, $H_{t}^{2}$, $H_{t}^{3}$, and
$H_{t}^{4}$. Then we can write the sum $D_{t+1}$ as
\begin{equation}\label{APL03}
  D_{t+1} = 4\,D_t + \Delta_t\,,
\end{equation}
where $\Delta_t$ is the sum of length over all shortest paths whose
end points are not in the same $H_{t}$ branch.

The paths that contribute to $\Delta_t$ must all go through at least
one of the four \emph{connecting nodes} (i.e., $W$, $X$, $Y$ and $Z$
in Fig.~\ref{copy}) at which the different $H_t$ branches are
connected. The analytical expression for $\Delta_t$, called the
length of crossing paths, is found below.

Denote $\Delta_t^{\alpha,\beta}$ as the sum of length for all
shortest paths with end points in $H_t^{\alpha}$ and $H_t^{\beta}$,
respectively. If $H_t^{\alpha}$ and $H_t^{\beta}$ meet at a
connecting node, $\Delta_t^{\alpha,\beta}$ rules out the paths where
either end point is that shared connecting node. For example, each
path contributed to $\Delta_t^{1,2}$ should not end at node $W$. If
$H_t^{\alpha}$ and $H_t^{\beta}$ do not meet,
$\Delta_t^{\alpha,\beta}$ excludes the paths where either end point
is any connecting node. For instance, each path contributed to
$\Delta_t^{1,3}$ should not end at nodes $W$, $X$, $Y$ or $Z$. Then
the total sum $\Delta_t$ is
\begin{align}
\Delta_t =& \,\Delta_t^{1,2} + \Delta_t^{1,3} + \Delta_t^{1,4}+
\Delta_t^{2,3} + \Delta_t^{2,4}+\Delta_t^{3,4}-4. \label{APL04}
\end{align}
The last term at the end compensates for the overcounting of certain
paths: the shortest path between $W$ and $Y$, with length $2$, is
included in $\Delta_t^{1,4}$ and $\Delta_t^{2,3}$; the shortest path
between $X$ and $Z$, also with length $2$, is included in both
$\Delta_t^{1,2}$ and $\Delta_t^{3,4}$.

By symmetry, $\Delta_t^{1,2} = \Delta_t^{1,4}=\Delta_t^{2,3} =
\Delta_t^{3,4}$ and $\Delta_t^{1,3} = \Delta_t^{2,4}$, so that
\begin{equation}\label{APL05}
\Delta_t = 4 \Delta_t^{1,2} + 2 \Delta_t^{1,3}-4.
\end{equation}
In order to find $\Delta_t^{1,2}$ and $\Delta_t^{1,3}$, we define
quantity $s_t$ as
\begin{equation}
s_t = \sum_{\substack{i \in H_t \\ i \ne W}} d_{iW}\,. \label{APL06}
\end{equation}

Considering the self-similar network structure, we can easily know
that at time $t+1$, the quantity $s_{t+1}$ evolves recursively as
\begin{eqnarray}
s_{t+1}
&=&2\,s_t+\left[s_t+(N_t-1)\right]+\left[s_t+(N_t-1)-2\right]\nonumber\\
&=&4\,s_t+\frac{4}{3}(4^t-1).\label{APL07}
\end{eqnarray}
Using $s_1=4$, we have
\begin{eqnarray}
s_t= \frac{1}{9} \left(4+5\times4^t+3\,t\times4^t\right).
\end{eqnarray}

On the other hand, by definition given above, we have
\begin{eqnarray}
  \Delta_t^{1,2} &=& \sum_{\substack{i \in H_t^{1},\,\,j\in
      H_t^{2}\\ i,j \ne W}} d_{ij}\nonumber\\
  &=& \sum_{\substack{i \in H_t^{1},\,\,j\in
      H_t^{2}\\ i,j \ne W}} (d_{iW} + d_{jW}) \nonumber\\
  &=& (N_t-1)\sum_{\substack{i \in H_t^{1}\\ i \ne W}} d_{iW} + (N_t-1) \sum_{\substack{j \in H_t^{2}\\ j \ne W}} d_{jW} \nonumber\\
  &=& 2(N_t-1)\sum_{\substack{i \in H_t^{1} \\ i \ne W}} d_{iW} \nonumber\\
  &=& 2(N_t-1)\,s_t.
\label{APL08}
\end{eqnarray}

Continue analogously,
\begin{eqnarray}
  \Delta_t^{1,3} &=& \sum_{\substack{i \in H_t^{1},\,i \ne W, Z \\ j\in
      H_t^{3}, j \ne X, Y}} d_{ij}\nonumber\\
  &=& \sum_{\substack{i \in H_t^{1},\,i \ne W, Z \\ d_{iW}<d_{iZ} \\ j\in
      H_t^{3}, j \ne X, Y}} (d_{iW}+d_{WX}+d_{jX})\nonumber\\
&\quad&+\sum_{\substack{i \in H_t^{1},\,i \ne W, Z \\ d_{iZ}<d_{iW} \\
j\in
      H_t^{3}, j \ne X, Y}} (d_{iZ}+d_{ZY}+d_{jY})\nonumber\\
  &=& 2\,\sum_{\substack{i \in H_t^{1},\,i \ne W, Z \\ d_{iW}<d_{iZ} \\ j\in
      H_t^{3}, j \ne X, Y}} (d_{iW}+d_{WX}+d_{jX})\,,
  \label{APL09}
\end{eqnarray}
where the symmetry property has been used. After simple
calculations, we obtain
\begin{eqnarray}
  \Delta_t^{1,3} =2(N_t-1)\,s_t&+&(N_t-1)^2-2\left(\frac{N_t}{2}\right)^2\nonumber\\&-&2(s_t+N_t-3)-1.
  \label{APL10}
\end{eqnarray}
Substituting Eqs.~(\ref{APL08}) and (\ref{APL10}) into
Eq.~(\ref{APL05}), we obtain after simplification
\begin{equation}\label{APL11}
\Delta_t =\frac{4}{9} \left(-2 + 11\times16^t +6\,t\times
16^{t}\right).
\end{equation}
Thus
\begin{equation}\label{APL12}
D_{ t +1} = 4\,D_t +\frac{4}{9} \left(-2 + 11\times16^t +6\,t\times
16^{t}\right).
\end{equation}
Using $D_1 =8$, Eq.~(\ref{APL12}) is solved inductively,
\begin{equation}\label{APL13}
 D_t = \frac{1}{27}\left(8 + 16\times4^{t} + 3\times16^t + 6t\times16^{t} \right).
\end{equation}
Inserting Eq.~(\ref{APL13}) into Eq.~(\ref{APL01}), one can obtain
the analytical expression for $d_t$:
\begin{equation}\label{APL14}
d_t = \frac{2}{3}\times\frac{8 + 16\times4^{t} + 3\times16^t +
6t\times16^{t}}{4 \times 16^t+10\times4^{t}+4},
\end{equation}
which approximates $t$ in the infinite $t$, implying that the APL
shows a logarithmic scaling with network order. Therefore, in the
case of $q=1$, the network exhibits a small-world behavior. We have
checked our analytic result against numerical calculations for
different network order up to $t=10$ which corresponds to $N_{10}=1
048 577$. In all the cases we obtain a complete agreement between
our theoretical formula and the results of numerical investigation.

\subsubsection{Case of $q=0$}

For this particular case, our networks turn out to be the
hierarchical lattice introduced in~\cite{BeOs79}, which is also
self-similar. Using a method similar to but a little different from
that applied in preceding subsection, we can compute analytically
the average path length $d_t$. We omit the calculation process and
give only the exact expression as below:
\begin{equation}\label{APL15}
d_t = \frac{22\times 2^t\times16^{t} +8^{t}(21t+42) + 27\times4^{t}
+ 98\times 2^t}{42 \times 16^t+105\times4^{t}+42}.
\end{equation}
Equation~(\ref{APL15}) recovers the previously obtained result
in~\cite{HiBe06} and has been confirmed by extensive numerical
simulations. In the large $t$ limit, $d_{t} \sim \frac{11}{21}2^t$.
On the other hand, for large $t$ limit, $N_t \sim 4^t$, so $d_{t}$
grows as a square root of the number of network nodes. Thus, in the
case of $q=0$, the network exhibits a `large-world' behavior of
typical node-node distances.

\subsubsection{Case of $0<q<1$}

\begin{figure}
\begin{center}
\includegraphics[width=.45\linewidth,trim=100 30 100 10]{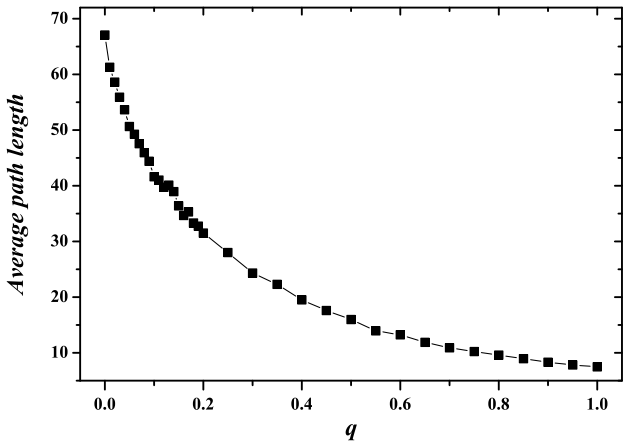}
\end{center}
\caption[kurzform]{\label{fig6} Graph of the dependence of the
average path length on the tunable parameter $q$.}
\end{figure}

For $0<q<1$, in order to obtain the variation of the average path
length with the parameter $q$, we have performed extensive numerical
simulations for different $q$ between 0 and 1. Simulations were
performed for network $N_{7}$ with order $10924$, averaging over 20
network samples for each value of $q$. In Fig.~\ref{fig6}, we plot
the average path length as a function of $q$. We observe that, when
$q$ increases from 0 to 1, the average path length drops drastically
from a very high value to a small one, which shows that there is a
crossover between small-world and `large-world'. This behavior is
similar to that in the WS model~\cite{WaSt98}.

%%%%%%%%%%%%%%%%%%%%%%%%%%%%%%%%%%%%%%%%%%%%%%%%%%%%%%%%%
% Figure 5
%%%%%%%%%%%%%%%%%%%%%%%%%%%%%%%%%%%%%%%%%%%%%%%%%%%%%%%%%%
%%%%%%%%%%%%%%%%%%%%%%%%%%%%%%%%%%%%%%%%%%%%%%%%%%%%%%%%%%

\section{Threshold of bond percolation}

As discussed in previous section, the networks exhibit many
interesting properties, i.e., they have the same degree sequence
independent of parameter $q$; they are scale-free and non-clustered;
and they display a crossover between ``large-world" and small-world.
All these features are not shared simultaneously by any previously
reported networks. Hence, it is worthwhile to investigate the
processes taking place upon the model to find the different impact
on dynamic precesses compared with other networks such as the BA
network. In what follows we will study bond percolation, which is
one of the most important issues in statistical
physics~\cite{StAh92}.

In bond percolation every bond (link or edge) on a specified graph
is independently either ``occupied'' with probability $\lambda$, or
not with the complementary probability $1-\lambda$. In our case the
percolation problem can be solved using the real-space
renormalization group technique~\cite{Mi75,Mi76,Ka76,Do03,RoAv07},
giving exact solution for the interesting quantity of percolation
threshold. Let us describe the procedure in application to the
network considered. Assuming that the network growth stops at a time
step $t\rightarrow \infty$, when the network is spoiled in the
following way: for a link present in the undamaged network, with the
probability $\lambda$ we retain it in the damaged network. Then we
invert the transformation in Fig. 1 and define $n=t- \tau$ for this
inverted transformation, which is actually a decimation
procedure~\cite{Do03}. Further, we introduce the probability
$\lambda_n$ that if two nodes are connected in the undamaged network
at $\tau =t-n$, then at the $n$th step of the decimation for the
damaged network, there exists a path between these vertices. Here,
$\lambda_0= \lambda$. We can easily obtain the following recursion
relation for $\lambda_n$
\begin{equation}\label{lambda01}
\lambda_{n+1}=q\,(\lambda_n+\lambda_n^3-\lambda_n^4)+(1-q)(2\lambda_n^2-\lambda_n^4).
\end{equation}

Equation~(\ref{lambda01}) has four roots (i.e., fixed points), among
which the root $\lambda=-\frac{1-q}{2}-\frac{1}{2}\sqrt{5-6q+q^2}$
is invalid, because it is less than 0. The other three fixed points
are as follows: two stable fixed points at $\lambda=0$ and
$\lambda=1$, and an unstable fixed point at $\lambda_c$ that is the
percolation threshold. The reason for the unstable fixed point
corresponding to the threshold is as follows: at any $0<\lambda_0<
\lambda_c$, $\lambda_n$ approaches 0 as $n\rightarrow \infty$, which
means there is no percolation; while at any $\lambda_c<\lambda_0<1$,
$\lambda_n$ approach 1, indicating an existence of the percolating
cluster.

\begin{figure}[h]
\centering\includegraphics[width=.30\linewidth,trim=110 20 110
20]{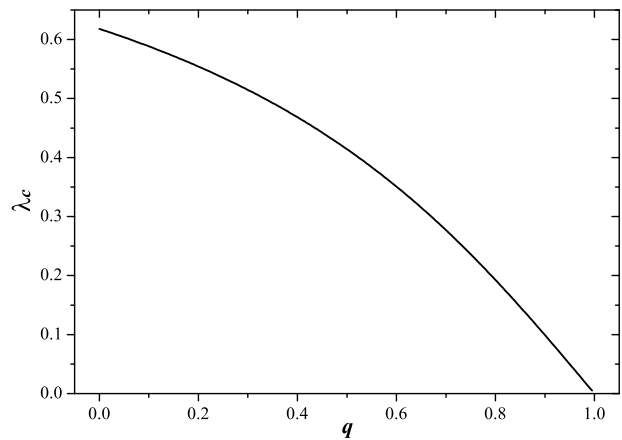} \caption{ The dependence relation of percolation
threshold $\lambda_c$ on the parameter $q$. }\label{fig4}
\end{figure}

The exact expression of $\lambda_c$ as a function of $q$ is
\begin{equation}\label{lambda02}
\lambda_c=-\frac{1-q}{2}+\frac{1}{2}\sqrt{5-6q+q^2}.
\end{equation}
Interestingly, for the case of $q=0$, $\lambda_c$ is equal to
$\frac{\sqrt{5}-1}{2}$, which is the inverse of the golden ratio
$\phi$ ($\phi=\frac{\sqrt{5}+1}{2}$) and is the same value as the
site percolation threshold for the ``B" lattice discussed
in~\cite{Sc06,Zf06}. We present the dependence of $\lambda_c$ on $q$
in Fig.~\ref{fig4}, which indicates that the threshold $\lambda_c$
decreases as $q$ increases. When $q$ grows from 0 to 1, $\lambda_c$
decreases from $\frac{\sqrt{5}-1}{2}\approx 0.618$ to 0.

%%%%%%%%%%%%%%%%%%%%%%%%%%%%%%%%%%%%%%%%%%%%%%%%%%%%%%%%%
% Figure 6
%%%%%%%%%%%%%%%%%%%%%%%%%%%%%%%%%%%%%%%%%%%%%%%%%%%%%%%%%%
%%%%%%%%%%%%%%%%%%%%%%%%%%%%%%%%%%%%%%%%%%%%%%%%%%%%%%%%%%

Thus, in a large range of parameter $q$ (i.e., $q<1$), there exists
a critical non-zero percolation threshold $\lambda_c$ such that for
$\lambda >\lambda_c$ a giant component appears spanning the entire
network, for $\lambda <\lambda_c$ there are only isolated small
clusters. The existence of percolation thresholds in our networks is
in sharp contrast with the null threshold found in a wide range of
previously studied scale-free
networks~\cite{CoErAvHa00,CaNeStWa00,BoVePa03,VaMo03,SeBo06}.

Note that since the susceptible-infected-removed (SIR) model can be
mapped to the bond percolation problem~\cite{MoNe00,Ne02,BoVePa03},
for the SIR model on our networks the epidemic prevalence undergoes
a phase transition at a finite threshold $\lambda_c$ of the
transmission probability. If infection rate is above $\lambda_c$,
the disease spreads and infects a finite fraction of the population.
On the contrary, when infection rate is below $\lambda_c$, the total
number of infected individuals is infinitesimally small in the limit
of very large populations. The existence of epidemic thresholds in
the present networks is compared to the result for some other
scale-free networks, where arbitrarily small infection rate shows
finite prevalence~\cite{MoPaVe02}.

From Eq.~(\ref{lambda02}), one can see that for different $q$, the
networks have distinct percolation thresholds. As known from
preceding section, the whole class of the networks exhibits
identical degree sequence (power-law degree distribution) and (zero)
clustering coefficient, which shows that degree distribution and
clustering coefficient are not sufficient to characterize the
threshold of bond percolation in scale-free networks. One may ask
why the considered networks have disparate percolation thresholds.
We speculate that the diverse thresholds in our networks lie with
the average path length, which needs further confirmation in the
future.

\section{Conclusions}

We have demonstrated that power-law degree distribution alone does
not suffice to characterize the percolation threshold on scale-free
networks under bond percolation. To this end, by introducing a
parameter $q$, we have presented a family of scale-free networks
with the same degree sequence and (zero) clustering coefficient. We
provided a detailed analysis of the topological features and showed
that the model exhibits a rich structural behavior. In particular,
using a renormalization method, we have derived an exact analytic
expression for the thresholds of bond percolation in our networks.
We found that finite thresholds are recovered for our networks in
the case of $q<1$, which is in contrast to the conventional wisdom
that null percolation threshold is an intrinsic nature of scale-free
networks. Therefore, care should be needed when making general
statements about the percolation problem in scale-free networks.

It should be mentioned that the model generation of scale-free
networks with the same degree sequence is a very common problem in
complex network research. Actually, in the study of the impacts of
other characteristics (besides degree distribution) of scale-free
networks on the dynamical processes defined on the networks, the
interference of power-law degree distribution should be avoided. In
this case, such a model is necessitated. Traditionally, the
interchanging algorithm (through rewiring two links between four end
points) is frequently used to achieve this goal~\cite{MaSn02}. But
this algorithm may lead to disconnection of the whole network. We
have shown that the scale-free networks proposed here have identical
degree sequence and are always connected. So our networks can
overcome above deficiency. They may be helpful for investigating how
other features (say, average path length), other than power-law
degree distribution, are relevant to the performance of scale-free
networks.

Finally, we stress that since we were only concerned with the
percolation phase transition point, we merely gave the exact
position of the percolation thresholds, omitting some other
properties of bond percolation, such as the value of the critical
exponents governing behavior close to the transition, the complete
distribution of the cluster sizes, and closed-form expressions for
the mean and variance of the distribution. All these are worth
studying further in the future, which is beyond the scope of this
paper.

\emph{Note added.}---A relevant publication~\cite{WaSaSo02} about
bond percolation has come to our attention, where the authors showed
that different percolation thresholds exist for different networks
having the same degree distribution (not degree sequence as
addressed in this present paper).

\begin{acknowledgments}

We would like to thank Yichao Zhang for support. This research was
supported by the National Basic Research Program of China under
Grant No. 2007CB310806, the National Natural Science Foundation of
China under Grant Nos. 60704044, 60873040 and 60873070, Shanghai
Leading Academic Discipline Project No. B114, and the Program for
New Century Excellent Talents in University of China (NCET-06-0376).
We thank a referee for informing us about Ref.~\cite{WaSaSo02}.

\end{acknowledgments}


\begin{thebibliography}{}

\bibitem{StAh92}
D. Stauffer and A. Aharony, \emph{Introduction to Percolation
Theory}, 2nd ed. (Taylor and Francis, London, 1992).

\bibitem{AlBa02} R. Albert and A.-L. Barab\'asi,
      %Statistical mechanics of complex networks,
       Rev. Mod. Phys. {\bf 74}, 47 (2002).

\bibitem{DoGoMe08} S. N. Dorogovtsev, A. V. Goltsev and J. F. F. Mendes,
      %Critical phenomena in complex networks,
       Rev. Mod. Phys. {\bf 80}, 1276 (2008).

\bibitem{AlJeBa00}
R. Albert, H. Jeong,  A.-L. Barab\'asi, Nature (London) {\bf 406},
378 (2000).

\bibitem{CaNeStWa00}
D. S. Callaway, M. E. J. Newman,  S. H. Strogatz, and D. J. Watts,
%   Network Robustness and Fragility: Percolation on Random Graphs
Phys. Rev. Lett. {\bf 85}, 5468 (2000).

\bibitem{CoErAvHa00}
R. Cohen, K. Erez, D. ben-Avraham,  S. Havlin, Phys. Rev. Lett. {\bf
85}, 4626 (2000).

\bibitem{CoErAvHa01}
R. Cohen, K. Erez, D. ben-Avraham,  S. Havlin, Phys. Rev. Lett. {\bf
86}, 3682 (2001);

\bibitem{MoNe00}
C. Moore and M. E. J. Newman, Phys. Rev. E {\bf 61}, 5678 (2000).

\bibitem{Ne02}
M. E. J. Newman, Phys. Rev. E {\bf 66}, 016128 (2002).

\bibitem{BoVePa03}
M. Bogu{\~n}{\'a}, R. Pastor-Satorras and A. Vespignani, Phys. Rev.
Lett. {\bf 90}, 028701 (2003).

\bibitem{Da05}
L. Dall'Asta, J. Stat. Mech.: Theory Exp. {\bf P08011} (2005).

\bibitem{ZhZhZoCh08}
Z. Z. Zhang, S. G. Zhou, T. Zou, and G. S. Chen, J. Stat. Mech.:
Theory Exp. {\bf P09008} (2008).

\bibitem{ScZf08}
C. R. Scullard and R. M. Ziff, Phys. Rev. Lett. {\bf 100}, 185701
(2008).

\bibitem{BrHa57}
S. R. Broadbent and J. M. Hammersley, Proc. Cambridge Philos. Soc.
{\bf 53}, 629 (1957).

\bibitem{Fl41}
P. J. Flory, J. Am. Chem. Soc. {\bf 63}, 3083 (1941).

\bibitem{Sc06}
C. R. Scullard, Phys. Rev. E {\bf 73}, 016107 (2006).

\bibitem{Zf06}
R. M. Ziff, Phys. Rev. E {\bf 73}, 016134 (2006).

\bibitem{Pi06}
W. Pietsch, Phys. Rev. E {\bf 73}, 066112 (2006).

\bibitem{BaAl99} A.-L. Barab\'asi and R. Albert,
      %Emergence of scaling in random networks,
       Science {\bf 286}, 509 (1999).

\bibitem{NeZi01}
M. E. J. Newman and R. M. Ziff, Phys. Rev. E {\bf 64}, 016706
(2001).

\bibitem{VaMo03}
A. V\'azquez and Y. Moreno, Phys. Rev. E 67, 015101(R) (2003).

\bibitem{SeBo06}
M. \'A. Serrano and M. Bogu{\~n}{\'a}, Phys. Rev. Lett. {\bf 97},
088701 (2006).

\bibitem{Ne03} M. E. J. Newman,
%The structure and function of complex networks,
SIAM Rev. {\bf 45}, 167 (2003).



\bibitem{WaSt98} D. J. Watts and H. Strogatz,
       %Collective dynamics of `small-world' networks,
        Nature (London) {\bf 393}, 440 (1998).

\bibitem{RoHaAv07}
H. D. Rozenfeld, S. Havlin, and D. ben-Avraham, New J. Phys. {\bf
9}, 175 (2007).

\bibitem{BeOs79}
A. N. Berker and S. Ostlund, J. Phys. C {\bf 12}, 4961 (1979).

\bibitem{HiBe06}
M. Hinczewski and A. N. Berker, Phys. Rev. E {\bf 73}, 066126
(2006).

\bibitem{Mi75}
A. A. Migdal, Zh. Eksp. Teor. Fiz. {\bf 69}, 1457 (1975).

\bibitem{Mi76}
A. A. Migdal, Sov. Phys. JETP {\bf 42}, 743 (1976).

\bibitem{Ka76}
L. P. Kadanoff, Ann. Phys. (N.Y.) {\bf 100}, 359 (1976).

\bibitem{Do03}
S. N. Dorogovtsev, Phys. Rev. E {\bf 67}, 045102(R) (2003).

\bibitem{RoAv07}
H. D. Rozenfeld and D. ben-Avraham, Phys. Rev. E {\bf 75}, 061102
(2007).

\bibitem{MoPaVe02}
Y. Moreno, R. Pastor-Satorrasand, and  A. Vespignani, Eur. Phys. J.
B {\bf 26},  521 (2002).

\bibitem{MaSn02}
S. Maslov, K. Sneppen, Science {\bf 296}, 910 (2002).

\bibitem{WaSaSo02}
C. P. Warren, L. M. Sander, and I. M. Sokolov, Phys. Rev. E {\bf
66}, 056105 (2002).


\end{thebibliography}
\end{document}